\documentclass[journal,draftclsnofoot,onecolumn,twoside,12pt]{IEEEtran}

\usepackage{fixltx2e}
\usepackage[cmex10]{amsmath}
\usepackage{amssymb}
\usepackage{algorithmic}
\usepackage{array}
\usepackage{mdwmath}
\usepackage{mdwtab}
\usepackage{cite}
\usepackage{url}
\usepackage{color}
\usepackage{xcolor}
\usepackage{gensymb}
\usepackage{textcomp}

\ifCLASSINFOpdf
  \usepackage[pdftex]{graphicx}
  \graphicspath{{./input/}}
  \DeclareGraphicsExtensions{.pdf,.png,.jpg,.jpeg}
\else
  \usepackage[dvips]{graphicx}
  \graphicspath{{./input/}}
  \DeclareGraphicsExtensions{.eps}
\fi
\usepackage{pgfplots}
\pgfplotsset{compat=newest}
\pgfplotsset{/tikz/every mark/.append style={solid}}
\usepackage[font=footnotesize,caption=false]{subfig}

\usepgfplotslibrary{external}
\renewcommand{\baselinestretch}{1.7}

\begin{document}

\newcommand{\hblack}[1]{\textcolor{black}{#1}}
\newcommand{\hred}[1]{\textcolor{red}{#1}}
\newcommand{\hblue}[1]{\textcolor{blue}{#1}}
\newcommand{\hgreen}[1]{\textcolor{green}{#1}}

\newcommand{\mc}[3]{\multicolumn{#1}{#2}{#3}}
\newcommand{\mr}[3]{\multirow{#1}{#2}{#3}}

\newcommand{\chalab}[1]{\label{cha:#1}}
\newcommand{\seclab}[1]{\label{sec:#1}}
\newcommand{\applab}[1]{\label{app:#1}}
\newcommand{\figlab}[1]{\label{fig:#1}}
\newcommand{\tablab}[1]{\label{tab:#1}}
\newcommand{\eqnlab}[1]{\label{eqn:#1}}

\newcommand{\charef}[1]{Chapter~\ref{cha:#1}}
\newcommand{\secref}[1]{Section~\ref{sec:#1}}
\newcommand{\appref}[1]{Appendix~\ref{app:#1}}
\newcommand{\figref}[1]{Fig.~\ref{fig:#1}}
\newcommand{\tabref}[1]{Table~\ref{tab:#1}}
\newcommand{\eqnref}[1]{(\ref{eqn:#1})}

\newcommand{\sizecorr}[1]{\makebox[0cm]{\phantom{$\displaystyle #1$}}}

\newcommand{\op}[1]{\operatorname{#1}}
\newcommand{\vm}[1]{\mathbf{#1}}
\newcommand{\set}[1]{\mathcal{#1}}
\newcommand{\ie}{,\ i.e.,\ }
\newcommand{\eg}{,\ e.g.,\ }
\newcommand{\dB}{\text{dB}}
\newcommand{\F}[3]{\operatorname{F}_{#1 \rightarrow #2}\left\{ #3 \right\}}
\newcommand{\Fi}[3]{\operatorname{F}^{-1}_{#1 \rightarrow #2}\left\lbrace #3 \right\rbrace}
\newcommand{\E}[1]{\operatorname{E}\left\lbrace #1 \right\rbrace}
\newcommand{\Earg}[2]{\operatorname{E}_{#1}\left\lbrace #2 \right\rbrace}
\newcommand{\var}[1]{\operatorname{var}\left\lbrace #1 \right\rbrace}
\newcommand{\cov}[2]{\operatorname{cov}\left\lbrace #1,#2 \right\rbrace}
\newcommand{\Prob}[1]{\operatorname{P}\left( #1 \right)}
\newcommand{\minim}[2]{\min_{#1}\left\lbrace #2 \right\rbrace}
\newcommand{\maxim}[2]{\max_{#1}\left\lbrace #2 \right\rbrace}
\newcommand{\vect}[1]{\operatorname{vec}\left\lbrace #1 \right\rbrace}
\newcommand{\unvect}[1]{\operatorname{unvec}\left\lbrace #1 \right\rbrace}
\newcommand{\Ndist}[1]{\mathcal{N}\left( #1 \right)}
\newcommand{\CNdist}[1]{\mathcal{CN}\left( #1 \right)}
\newcommand{\tr}[1]{\operatorname{tr}\left\lbrace #1 \right\rbrace}
\newcommand{\diag}[1]{\operatorname{diag}\left\lbrace #1 \right\rbrace}
\newcommand{\re}[1]{\Re \left\lbrace #1 \right\rbrace}
\newcommand{\im}[1]{\Im \left\lbrace #1 \right\rbrace}
\newcommand{\sinc}[1]{\operatorname{sinc}\left( #1 \right)}
\newcommand{\logmod}[1]{\log \left( #1 \right)}
\newcommand{\logdet}[1]{\log \operatorname{det}\left( #1 \right)}

\newcommand{\opH}{\operatorname{H}}
\newcommand{\IBF}{I_{\text{BF}}}
\newcommand{\IBFa}{I_{\text{BF,a}}}

\definecolor{lightblue}{rgb}{0.2,0.6,1}
\definecolor{darkgreen}{rgb}{0.2,0.7,0.2}
\pgfplotscreateplotcyclelist{mylist}{
 {black},
 {lightblue},
 {red},
 {darkgreen},
}
\pgfplotscreateplotcyclelist{mylist_markers}{
 {black,mark=+},
 {lightblue,mark=o},
 {red,mark=x},
 {darkgreen,mark=diamond},
}
\pgfplotscreateplotcyclelist{mylist_approx}{
 {black},
 {black, dashed},
 {lightblue},
 {lightblue, dashed},
 {red},
 {red, dashed},
 {darkgreen},
 {darkgreen, dashed},
}
\pgfplotscreateplotcyclelist{mylist_LQS_SISO}{
 {black, line width=1pt},
 {black, dashed, line width=1pt},
 {black, dotted, line width=1pt},
 {lightblue, line width=1pt},
 {lightblue, dashed, line width=1pt},
 {lightblue, dotted, line width=1pt},
}
\pgfplotscreateplotcyclelist{mylist_LQS_MIMO}{
 {black, line width=1pt},
 {black, dashed, line width=1pt},
 {lightblue, line width=1pt}, 
 {lightblue, dashed, line width=1pt},
 {red, line width=1pt},
}

\bstctlcite{BSTcontrol}

\title{Analysis of the Local Quasi-Stationarity of Measured Dual-Polarized MIMO Channels}

\author{Adrian~Ispas,~Christian~Schneider,~Gerd~Ascheid\IEEEmembership{,~Senior~Member,~IEEE},~and~Reiner~Thom\"a\IEEEmembership{,~Fellow,~IEEE}%
\thanks{This work was supported by the Ultra high-speed Mobile Information and Communication (UMIC) research centre and the European Commission in the framework of the FP7 Network of Excellence in Wireless COMmunications NEWCOM++. Parts of this work were presented in \cite{Ispas_LQSRegions, Ispas_LQSRegions_EB}.}%
\thanks{Adrian Ispas was with the Chair for Integrated Signal Processing Systems, RWTH Aachen University. He is now with Rohde \& Schwarz GmbH \& Co. KG, 81671 Munich, Germany (e-mail: ispas@iss.rwth-aachen.de).}
\thanks{Gerd Ascheid is with the Chair for Integrated Signal Processing Systems, RWTH Aachen University, 52056 Aachen, Germany (e-mail: ascheid@iss.rwth-aachen.de).}%
\thanks{Christian Schneider and Reiner Thom\"a are with the Institute for Information Technology, Ilmenau University of Technology, 98684 Ilmenau, Germany (e-mail: \{christian.schneider, reiner.thomae\}@tu-ilmenau.de).}}




\maketitle

\vspace{-1cm}
\begin{abstract}
It is common practice in wireless communications to assume strict or wide-sense stationarity of the wireless channel in time and frequency. While this approximation has some physical justification, it is only valid inside certain time-frequency regions. This paper presents an elaborate characterization of the non-stationarity of wireless dual-polarized channels in time. The evaluation is based on urban macrocell measurements performed at $2.53$~GHz. In order to define local quasi-stationarity (LQS) regions\ie regions in which the change of certain channel statistics is deemed insignificant, we resort to the performance degradation of selected algorithms specific to channel estimation and beamforming. Additionally, we compare our results to commonly used measures in the literature. We find that the polarization, the antenna spacing, and the opening angle of the antennas into the propagation channel can strongly influence the non-stationarity of the observed channel. The obtained LQS regions can be of significant size\ie several meters, and thus the reuse of channel statistics over large distances is meaningful (in an average sense) for certain algorithms. Furthermore, we conclude that, from a system perspective, a proper non-stationarity analysis should be based on the considered algorithm.
\end{abstract}


\section{Introduction}

\IEEEPARstart{I}{n} wireless communications, it is common to assume the wireless channel to be a randomly time-variant linear channel. Besides the assumption on linearity, the random process describing the channel is mostly assumed to follow strict or wide-sense stationarity in time and frequency. A wide-sense stationary channel in time/frequency has constant first- and second-order statistics over time/frequency. However, the statistics of realistic channels change over time and frequency. Statistical signal processing algorithms that rely on knowledge of second-order statistics of the channel thus have to update this knowledge when their performance is degraded due to the statistical mismatch. Therefore, it is crucial to assess the size of the time-frequency regions inside which the change of the channel statistics is not significant with respect to a certain performance measure.

A wide-sense stationary channel in time and frequency is equivalent to the wide-sense stationary and uncorrelated scattering (WSSUS) channel commonly encountered in the literature. Based on the WSSUS assumption, the so-called quasi-WSSUS channel model is introduced for wireless channels in \cite{Bello_RTVL_Channels}. It considers the fact that channel statistics change on a large scale and thus it separates the channel fluctuations into fast and slow variations. This results in local WSSUS channels\ie WSSUS channels valid in local time-frequency regions. In \cite{Matz_Non-WSSUS_Channels}, a framework for the stochastic treatment of non-WSSUS channels is proposed. The assumption of any form of stationarity is dropped, while it is highlighted that typical wireless channels are doubly underspread (DU)\ie underspread in dispersion and correlation. Dispersion underspread channels are well known in the context of WSSUS channels; they are characterized by a maximal effective\ie relevant, delay and Doppler frequency shift whose product is much smaller than one. The correlation underspread property arises in the context of non-stationary channels; it states that the product of the effective correlation length in delay and the effective correlation length in Doppler is much smaller than the product of the maximal effective delay and the maximal effective Doppler. The DU property implies that the time-frequency regions of approximately constant channel statistics are much larger than the time-frequency coherence regions of the channel. This fact yields substantial theoretical simplifications of practical importance, such as a proper definition of a time-dependent power spectral density (PSD) for non-stationary random processes.

Recently, multiple-input and multiple-output (MIMO) systems that exploit the polarization domain \cite{Andrews_Tripling_Capacity_Polarization, Poon_Degrees_of_Freedom_Polarization, Landmann_Polarization} have received increased attention. The main advantage is the high decorrelation that occurs over orthogonally polarized antennas. This allows for compact antenna array designs by making use of\eg co-located dual-polarized (DP) antennas. The choice of the polarization at the transmitter (TX) and the receiver (RX) can result in substantially different propagation conditions. Depending on the polarization combination, different multipath components of the channel will determine the propagation conditions. For an extensive literature overview of experimental results related to DP channels, we refer the reader to \cite{Oestges_DP_Model_and_System_Eval}. It is thus of interest to extend the non-stationarity analysis to DP channels. We expect the polarization to have a noticeable impact on the non-stationarity of the channel. However, other parameters like the antenna spacing or the opening angle of the antennas into the propagation channel will influence the non-stationarity of the channel as well.

Early empirical investigations of non-stationary wireless single-input and single-output (SISO) channels in time and frequency can be found in \cite{Dossi_Statistical_Analysis} for an indoor channel and in \cite{Gehring_Stationarity} for an outdoor channel. The non-stationarity is typically characterized by comparing selected channel statistics over time/frequency with certain measures. The choice of the measure is obviously critical in assessing the degree of non-stationarity. Typically, one resorts to measures based on the standard Euclidean inner product. For a MIMO channel, in \cite{Herdin_CMD_1} and \cite{Herdin_PhD_Thesis}, the by now very popular correlation matrix distance (CMD) is introduced. It characterizes only the dissimilarity of the spatial properties of the channel and neglects the time-frequency ones for simplicity. In the literature, various contributions study the degree of non-stationarity of the wireless channel, see \cite{Dossi_Statistical_Analysis, Gehring_Stationarity, Viering_Validity_Spatial_Cov, Herdin_CMD_1, Herdin_CMD_2, Herdin_PhD_Thesis, Renaudin_Non-Stationary_MIMO_Channels, Paier_Vehicular_Channel_LSF, Bernado_Vehicular_Channel_Spect_Div, Aldayel_Master_Thesis, Ispas_LQSRegions, Ispas_LQSRegions_EB}. However, an extensive and thorough measurement-based characterization of the non-stationarity of wireless channels is still lacking. Moreover, most contributions do not consider the impact of the polarization of the channel on the degree of non-stationarity.

The choice of a proper measure, used to determine the region in which the change of the channel statistics is deemed insignificant, depends on the considered wireless communication algorithm. The reason is that the measure defines which channel statistics are being considered and how they are being used. Typically, a subset of the second-order channel statistics is considered\eg only some spatial properties in the form of correlation matrices. Consider\eg \cite{Herdin_CMD_2}, where a MIMO prefiltering technique is used to simulatively compare the bit error rate to the CMD in an indoor scenario. This approach can be extended by assessing the degree of non-stationarity based on the performance degradation due to outdated knowledge of the channel statistics. One can then define a maximal performance degradation above which the change of the channel statistics is considered substantial and thus its knowledge has to be updated. We name the resulting regions local quasi-stationarity (LQS) regions. Such an approach has been exemplarily demonstrated in \cite{Ispas_Mismatched_Wiener_Filtering_Journal} for the mean square error (MSE) degradation of a channel estimation algorithm.

In \cite{Ispas_LQSRegions}, we performed an exemplary analysis of the degree of non-stationarity of the channel for the spatial domains and the joint delay-Doppler domain by studying LQS regions based on reference measures from literature. The work \cite{Ispas_LQSRegions_EB} extended the analysis of the degree of non-stationarity of the spatial domains by comparing the resulting LQS regions to those of an approach based on the performance degradation of a beamforming technique.

\textit{Contributions:} In this paper, we perform an elaborate analysis of the local quasi-stationarity of measured DP MIMO wireless channels in time. Our approach is connected to selected algorithms that are commonly used in wireless communications; as such our results are important to the operation of these algorithms. The analysis is based on urban macrocell measurements at $2.53$~GHz relevant to 3GPP Long Term Evolution (LTE). As wireless communication algorithms typically only use a subset of the channel statistics, we perform our LQS analysis in several different domains\ie Doppler, delay, and space. The resulting LQS regions in time are mapped to the traveled distance of the mobile terminal (MT); they thus reflect the partial non-stationarity of the channel over distance in different domains. Our detailed contributions are as follows:
\begin{itemize}
 \item We analyze the LQS regions in the delay and the Doppler domain for DP channels by comparing the results of the collinearity measure to the ones of an MSE-based measure of a channel estimator.
 \item We analyze the LQS regions in the spatial domain at the BS and the MT side, for several $4\times4$ and $2\times2$ MIMO setups that are single-polarized (SP) or DP. Here, we compare the results of the collinearity measure\ie the CMD, to the ones of a measure based on the signal-to-noise-ratio (SNR) of a beamforming algorithm. We further give an algorithmic interpretation of the CMD in terms of an SNR.
 \item The results reveal that LQS regions can be quite large\ie several meters long, and thus the reuse of channel statistics over a distance defined by the LQS regions is meaningful (in an average sense) for certain algorithms.
 \item We find that a proper analysis of the non-stationarity of the channel requires the use of a measure adapted to the specific purpose. From a system perspective, this measure should reflect the performance degradation of the considered algorithm due to the non-stationarity of the channel.
\end{itemize}

Compared to our previous works\cite{Ispas_LQSRegions, Ispas_LQSRegions_EB}, we provide a thorough and extensive measurement-based evaluation of DP MIMO channels covering the complete reference scenario of the measurement campaign. Here, the methodology to obtain the LQS regions is based on measures averaged over the whole reference scenario; thus, we obtain LQS regions that are representative for this scenario. Moreover, we compare the LQS distances using algorithmic measures and measures based on the standard Euclidean inner product for all domains\ie the delay, the Doppler, and the spatial domain at the BS and the MT side. We further characterize the standard deviation of the measures and the correlation between all measures.

\textit{Structure:} After introducing the basic characterization of non-stationary channels in \secref{non-stationary_channels}, we present the theoretical basis of the analysis of the non-stationarity of the wireless channel in \secref{non-stationary_analysis}. In \secref{LQS_channels}, we introduce the definition of LQS regions used in the present work. The measurement campaign, the considered antenna setups, and the processing of the measurement data are discussed in \secref{measurements_and_processing}. In \secref{results}, we show and evaluate the obtained results. Finally, in \secref{conclusion}, we conclude our work.

\textit{Notation:} The $n$-dimensional convolution of $x(\cdot)$ and $y(\cdot)$ is represented by $(x \ast_n y)(\cdot)$. The cardinality of the set $\mathcal{A}$ is denoted by $|\mathcal{A}|$. We use lowercase and uppercase boldface letters to designate vectors and matrices, respectively. For a matrix $\vm{A}$, the (element-wise) complex conjugate, the transpose, and the conjugate transpose are denoted by $\vm{A}^{\ast}$, $\vm{A}^T$, and $\vm{A}^H$, respectively. The trace of a square matrix $\vm{A}$ is written as $\tr{\vm{A}}$. For a matrix $\vm{A}$, $||\vm{A}||_F$ denotes the Frobenius norm. We use $[\vm{A}]_{k,l}$ to denote the element in the $k$-th row and the $l$-th column of $\vm{A}$. The expectation of a random variable $x$ is denoted by $\E{x}$. The imaginary unit is designated as $j$.

\section{Non-Stationary Channels}
\seclab{non-stationary_channels}

As the wireless radio channel is continuous in the time domain $t$ and the frequency domain $f$ by nature, we first present the baseband channel and the corresponding statistical parameters as such. In a second step, we derive a description of the statistical channel paraemeters based on a discretized channel in time and frequency. The resulting expression can be directly applied to measured channel data.

The local scattering function (LSF) is an extension of the scattering function in the context of WSSUS channels to the non-stationary case \cite{Matz_Non-WSSUS_Channels}. It is thus a time- and frequency-dependent PSD in the delay and Doppler domain. For a channel $\opH$, the LSF is defined as
\begin{IEEEeqnarray}{rCl}
 C_{\opH}(t,f;\nu,\tau) &=& \int_{-\infty}^{\infty} \int_{-\infty}^{\infty} R_{\opH}(t,f;\Delta_t,\Delta_f) e^{-j2\pi \left( \nu \Delta_t - \tau \Delta_f \right)} d\Delta_t d\Delta_f
\end{IEEEeqnarray}
with the Doppler shift $\nu$, the correlation function
\begin{IEEEeqnarray}{rCl}
 R_{\opH}(t,f;\Delta_t,\Delta_f) &=& \E{ L_{\opH}(t,f+\Delta_f) L_{\opH}^{\ast}(t-\Delta_t,f) }
\end{IEEEeqnarray}
and the time-varying transfer function
\begin{IEEEeqnarray}{rCl}
 L_{\opH}(t,f) &=& \int_{-\infty}^{\infty} h(t,\tau) e^{-j2\pi f \tau} d\tau .
\end{IEEEeqnarray}
A second-order stationary channel in time and frequency has a constant LSF over $t$ and $f$, respectively. Furthermore, note that, in the context of zero-mean random processes, second-order stationarity over time and frequency is equivalent to uncorrelatedness in Doppler and delay, respectively \cite{Bello_RTVL_Channels}.

\subsection{Doubly Underspread Channels}
\seclab{DU_channels}

In \cite{Matz_Non-WSSUS_Channels}, the class of DU channels is introduced. These channels are, on the one hand, \textit{dispersion} underspread with a maximal delay-Doppler product $\tau_{\text{max}} \nu_{\text{max}} \ll 1$. Here, $\tau_{\text{max}}$ and $\nu_{\text{max}}$ denote the maximal (effective) delay and Doppler, respectively. This property is well known from WSSUS channels. On the other hand, they are \textit{correlation} underspread with $\Delta_{\tau,\text{max}} \Delta_{\nu,\text{max}} / (\tau_{\text{max}} \nu_{\text{max}}) \ll 1$, where $\Delta_{\tau,\text{max}}$ and $\Delta_{\nu,\text{max}}$ denote the maximal (effective) correlation in delay and Doppler, respectively. This essentially means that the time-frequency coherence region of size $(\tau_{\text{max}} \nu_{\text{max}})^{-1}$ is much smaller than the time-frequency stationarity region of size $(\Delta_{\tau,\text{max}} \Delta_{\nu,\text{max}})^{-1}$. We refer to \cite{Matz_Non-WSSUS_Channels} for further details. While the dispersion underspread property arises in the context of WSSUS channels, the correlation underspread property is specific to non-stationary channels. For (zero-mean) WSSUS channels, different delay and Doppler components are uncorrelated per definition; therefore, the correlation underspread property is not meaningful in this context.

The LSF has some deficiencies\eg it is not guaranteed to be non-negative. For DU channels, it is possible to define generalized local scattering functions (GLSFs) \cite{Matz_Non-WSSUS_Channels}
\begin{IEEEeqnarray}{rCl}
 C_{\opH}^{(\Phi)} (t,f;\nu,\tau) &=& (C_{\opH} \ast_4 \Phi) (t,f;\nu,\tau)
 \eqnlab{GLSF}
\end{IEEEeqnarray}
with
\begin{IEEEeqnarray}{rCl}
 \Phi(t,f;\nu,\tau) &=& \sum_{s=1}^S \gamma_s \int\limits_{-\infty}^{\infty} \int\limits_{-\infty}^{\infty} L_{\op{G}_s}^{\ast}(-t,-f + \Delta_f) L_{\op{G}_s}(-t-\Delta_t,-f) e^{-j2\pi (\nu \Delta_t - \tau \Delta_f)} d \Delta_t d \Delta_f .\IEEEeqnarraynumspace
\end{IEEEeqnarray}
Here, $L_{\op{G}_s}(t,f)$ are windowing functions in time-frequency normalized to unit-energy, $\gamma_s \geq 0$ normalizing constants, and $S$ the number of used windows. From this, it can be seen that a GLSF is a smoothed version of the LSF. The GLSFs have practically important properties: they are real-valued and non-negative and, for DU channels, approximately equivalent. For further details, we refer to \cite{Matz_Non-WSSUS_Channels}. An alternate form of the GLSF can be obtained by rewriting \eqnref{GLSF} as
\begin{IEEEeqnarray}{rCl}
 C_{\opH}^{(\Phi)} (t,f;\nu,\tau) &=& \sum_{s=1}^{S} \gamma_s \E{ \left| H^{(\op{G}_s)}(t,f;\nu,\tau) \right|^2 }
 \eqnlab{GLSF_alt}
\end{IEEEeqnarray}
with
\begin{IEEEeqnarray}{rCl}
 H^{(\op{G}_s)}(t,f;\nu,\tau) &=& \int_{-\infty}^{\infty} \int_{-\infty}^{\infty} L_{\op{G}_s}^{\ast}(t'-t,f'-f) L_{\opH}(t',f') e^{-j2\pi \left( \nu t' - \tau f' \right)} dt' df'.
 \eqnlab{GLSF_alt_help}
\end{IEEEeqnarray}

In case we only want to study the Doppler or the delay properties, we can define the PSDs in Doppler and delay as
\begin{IEEEeqnarray}{rCl}
 C_{\nu}^{(\Phi)}(t,f;\nu) &=& \int_{-\infty}^{\infty} C_{\opH}^{(\Phi)} (t,f;\nu,\tau) d\tau
 \eqnlab{PSD_Doppler} \\
 C_{\tau}^{(\Phi)}(t,f;\tau) &=& \int_{-\infty}^{\infty} C_{\opH}^{(\Phi)} (t,f;\nu,\tau) d\nu
 \eqnlab{PSD_delay}
\end{IEEEeqnarray}
respectively.

\subsection{Non-Stationary MIMO Channels}
\seclab{non-stationary_MIMO_channels}

We are also interested in the analysis of non-stationary MIMO channels, therefore, we could extend the above definitions to the MIMO case. However, for MIMO channels, we focus only on the study of the channel statistics in the spatial domains. Furthermore, note that we only consider the case of non-stationarity in time and frequency\ie the non-WSSUS case, and do not consider stationarity over the TX or the RX antenna array \cite{Herdin_PhD_Thesis}. In order to study the time-frequency dependency of the channel statistics in space only, we can define the TX and the RX correlation matrices as
\begin{IEEEeqnarray}{rCl}
 \vm{R}_{\text{TX}}(t,f) &=& \E{ \vm{H}^T(t,f) \vm{H}^{\ast}(t,f) }
 \eqnlab{corr_matrix_TX}\\
 \vm{R}_{\text{RX}}(t,f) &=& \E{ \vm{H}(t,f) \vm{H}^H(t,f) }
 \eqnlab{corr_matrix_RX}
\end{IEEEeqnarray}
with the $N_{\text{RX}} \times N_{\text{TX}}$ channel matrix $\vm{H}(t,f)$. Here, $\left[ \vm{H}(t,f) \right]_{k,l}$ for $k=1,\ldots,N_{\text{RX}}$ and $l=1,\ldots,N_{\text{TX}}$ is the time-varying transfer function of the MIMO sub-link from the TX element $l$ to the RX element $k$. $N_{\text{TX}}$ and $N_{\text{RX}}$ denote the number of antennas at the TX and the RX side, respectively.

\section{Non-Stationarity Analysis}
\seclab{non-stationary_analysis}

In the previous section, we introduced the second-order moments of the channel we consider when analyzing the non-stationarity of the channel. Now, we describe the analysis of the non-stationarity based on these channel statistics using commonly used measures from literature \cite{Herdin_PhD_Thesis, Renaudin_Non-Stationary_MIMO_Channels, Paier_Vehicular_Channel_LSF} as well as measures that are more suitable from a system engineer's point of view. Note that all the measures presented here only characterize the non-stationarity based on second-order moments of the channel. For the remainder of this work, we generally omit the frequency argument since we do not study the non-stationarity of the channel in frequency.

\subsection{Measures Based on an Inner Product}
\seclab{measures_ip}

\subsubsection{Doppler and Delay Domain}

As a basis for the non-stationarity analysis in the Doppler and the delay domain, we use the standard inner product for square-integrable functions between two PSDs at different time instants as a measure. For the PSD in the Doppler domain, we have
\begin{IEEEeqnarray}{rCl}
 \int_{-\infty}^{\infty} C_{\nu}^{(\Phi)} (t,f;\nu) C_{\nu}^{(\Phi)} (t',f;\nu) d\nu .
 \eqnlab{inner_product_Doppler}
\end{IEEEeqnarray}
Due to the correlation underspread property of the DU condition, we have an effectively finite correlation in time and in frequency. The maximal effective correlation in time and in frequency is denoted by $\Delta_{t,\text{max}}$ and $\Delta_{f,\text{max}}$, respectively.\footnote{By effectively finite correlation in time and frequency, we mean that the autocorrelation function of the channel drops below a small threshold value outside a region defined by the lengths $2 \Delta_{t,\text{max}}$ and $2 \Delta_{f,\text{max}}$ in time and frequency, respectively.} Thus, using a sinc expansion, it can be shown that
\begin{IEEEeqnarray}{rCl}
 \!\!\!\!\!\!\!\! \int_{-\infty}^{\infty} C_{\nu}^{(\Phi)} \left(\frac{m}{B_{\nu}},\frac{q}{B_{\tau}};\nu\right) C_{\nu}^{(\Phi)} \left(\frac{m'}{B_{\nu}},\frac{q}{B_{\tau}};\nu\right) d\nu &\approx& \frac{1}{B_{\Delta_t}} \sum_{p=-\frac{B_p-1}{2}}^{\frac{B_p-1}{2}} C_{\nu}^{(\Phi)}[m,q;p] C_{\nu}^{(\Phi)}[m',q;p]
 \eqnlab{inner_product_Doppler_disc}
\end{IEEEeqnarray}
holds with $C_{\nu}^{(\Phi)}[m,q;p] = C_{\nu}^{(\Phi)} \left( \frac{m}{B_{\nu}},\frac{q}{B_{\tau}};\frac{p}{B_{\Delta_t}} \right)$, $B_{\nu} > 2\nu_{\text{max}}$, $B_{\tau} > \tau_{\text{max}}$, and $B_{\Delta_t} > 2 \Delta_{t,\text{max}}$, where $B_{\nu}$ and $B_{\Delta_t}$ are chosen such that $B_p = B_{\nu} B_{\Delta_t}$ is an odd integer. Based on \eqnref{inner_product_Doppler_disc}, we define the collinearity of the PSD in the Doppler domain between different time instants $m$ and $m'$:
\begin{IEEEeqnarray}{rCl}
 \eta_{\text{col},\vm{C}_{\nu}^{(\Phi)}}[m,m'] &=& \frac{\tr{{\vm{C}_{\nu}^{(\Phi)}}[m,q] \vm{C}_{\nu}^{(\Phi)}[m',q]}}{\left\| \vm{C}_{\nu}^{(\Phi)}[m,q] \right\|_F \left\| \vm{C}_{\nu}^{(\Phi)}[m',q] \right\|_F}
 \eqnlab{col_PSD_Doppler}
\end{IEEEeqnarray}
where the diagonal $B_p \times B_p$ matrices $\vm{C}_{\nu}^{(\Phi)}[m,q]$ are defined by $\left[\vm{C}_{\nu}^{(\Phi)}[m,q]\right]_{k,k} = C_{\nu}^{(\Phi)}[m,q;k-(B_p+1)/2]$ for $k=1,\ldots,B_p$. We can then similarly define $\eta_{\text{col},\vm{C}_{\tau}^{(\Phi)}}[m,m']$ with the diagonal $B_n \times B_n$ matrices $\vm{C}_{\tau}^{(\Phi)}[m,q]$ defined by $\left[\vm{C}_{\tau}^{(\Phi)}[m,q]\right]_{k,k} = C_{\tau}^{(\Phi)}[m,q;k-1]$ for $k = 1,\ldots,B_n$. Here, we have $C_{\tau}^{(\Phi)}[m,q;n] = C_{\tau}^{(\Phi)} \left( \frac{m}{B_{\nu}},\frac{q}{B_{\tau}};\frac{n}{B_{\Delta_f}} \right)$, $B_{\Delta_f} > 2 \Delta_{f,\text{max}}$, and the odd integer $B_n = B_{\tau} B_{\Delta_f}$.

\subsubsection{Spatial Domains}

In order to analyze only the spatial properties of the channel, we again consider the collinearity for the time instants $m$ and $m'$:
\begin{IEEEeqnarray}{rCl}
 \eta_{\text{col},\vm{R}_k} [m,m'] &=& \frac{\tr{\vm{R}_k[m,q] \vm{R}_k[m',q]}}{\left\| \vm{R}_k[m,q] \right\|_F \left\| \vm{R}_k[m',q] \right\|_F}
 \eqnlab{col_MIMO}
\end{IEEEeqnarray}
where the Hermitian and positive semidefinite matrix $\vm{R}_k[m,q]$ is either the TX correlation matrix $\vm{R}_{\text{TX}}[m,q] = \vm{R}_{\text{TX}}(m/B_{\nu},q/B_{\tau})$ or the RX correlation matrix $\vm{R}_{\text{RX}}[m,q] = \vm{R}_{\text{RX}}(m/B_{\nu},q/B_{\tau})$ for $m, q\in \mathbb{Z}$. Similarly, we have $\vm{H}[m,q] = \vm{H}(m/B_{\nu},q/B_{\tau})$. Therefore, with \eqnref{col_MIMO}, one can choose between analyzing the spatial properties at the TX or at the RX. The collinearity can be represented in terms of the CMD proposed in \cite{Herdin_CMD_1, Herdin_PhD_Thesis}:
\begin{IEEEeqnarray}{rCl}
 \text{CMD}_k[m,m'] &=& 1 - \eta_{\text{col},\vm{R}_k} [m,m']
 \eqnlab{cmd}
\end{IEEEeqnarray}
where $k$ stands for ``TX'' or ``RX''.

\subsection{Measures Based on an Algorithmic View}
\seclab{measures_alg}

We now introduce measures that relate the characterization of the non-stationarity to an algorithmic view. Therefore, the resulting non-stationarity analysis will not represent the pure channel anymore, but it will be connected to the chosen algorithm.

\subsubsection{Doppler and Delay Domain}

For the Doppler and the delay domain, we consider pilot-based channel estimation over time and frequency, respectively. Here, the pilot symbols are represented by the measurement samples. We start with the Doppler case\ie we consider estimation of a frequency-flat fading channel over time. Consider the received signal $y[m] = h[m] x[m] + n[m]$, where $\{h[m]\}$ is the proper complex zero-mean channel process, $\{x[m]\}$ is the transmitted sequence that consists of random data symbols and periodically inserted deterministic pilot symbols with power $\sigma_p^2$, and $\{n[m]\}$ is a zero-mean white proper Gaussian noise process with known variance $\sigma_n^2 > 0$. The processes $\{h[m]\}$, $\{x[m]\}$, and $\{n[m]\}$ are assumed to be mutually independent. Furthermore, we define the ratio $\gamma = \sigma_p^2 / \sigma_n^2$. From \cite{Ispas_Mismatched_Wiener_Filtering_Journal}, we have the MSE of the linear minimum MSE channel estimate in terms of the PSD of the underlying (zero-mean) random process. The mismatched MSE at time instant $m$ using statistical knowledge from time instant $m'$ with a pilot spacing $L$ and the interval length $N$ is given by $\tilde{\sigma}_{\nu,N,L}^2[m,m']$, see \cite{Ispas_Mismatched_Wiener_Filtering_Journal} for the corresponding derivation. The matched MSE at time instant $m$ follows as $\sigma_{\nu,N,L}^2[m] = \tilde{\sigma}_{\nu,N,L}^2[m,m]$. An approximate expression for the mismatched MSE is \cite{Ispas_Mismatched_Wiener_Filtering_Journal}:
\begin{IEEEeqnarray}{rCl}
 \tilde{\sigma}_{\nu,\text{ap},L}^2[m,m'] &=& \frac{1}{B_p} \sum_{p=-\frac{B_p-1}{2}}^{\frac{B_p-1}{2}} \frac{\gamma^{-2} B_{\nu} C_{\nu}^{(\Phi)}[m;p] + \gamma^{-1} B_{\nu}^2 \left( C_{\nu}^{(\Phi)}[m';p] \right)^2}{\left( B_{\nu} C_{\nu}^{(\Phi)}[m';p] + \gamma^{-1} \right)^2} .
 \eqnlab{mse_inf_fil_mismatch}
\end{IEEEeqnarray}
The approximate matched MSE at time instant $m$ is $\sigma_{\nu,\text{ap},L}^2[m] = \tilde{\sigma}_{\nu,\text{ap},L}^2[m,m]$.
For further details, we refer to \cite{Ispas_Mismatched_Wiener_Filtering_Journal}. In order to characterize the loss in MSE at time instant $m$ due to mismatched statistical knowledge from time instant $m'$, we define the relative MSE
\begin{IEEEeqnarray}{rCl}
 \eta_{\text{MSE},\nu,k,L}[m,m'] = \frac{\sigma_{\nu,k,L}^2[m]}{\tilde{\sigma}_{\nu,k,L}^2[m,m']}
 \eqnlab{mse_rel}
\end{IEEEeqnarray}
where $k$ is to be substituted by $N$ or ``ap''. The relative MSE in the delay case\ie considering estimation over frequency, follows analogoulsy as $\eta_{\text{MSE},\tau,k,L}[m,m']$.

\subsubsection{Spatial Domains}

For the spatial domains, we consider transmission over a frequency-flat fading channel. We use a simple strategy with a single transmitted stream based on the knowledge of the channel statistics at the TX. Specifically, we choose statistical transmit beamforming\ie linear rank-one precoding, with the precoding vector $\vm{u}_{\text{TX},\text{max}}^{\ast}[m]$. Here, $\vm{u}_{\text{TX},\text{max}}[m]$ is an eigenvector of the (frequency-independent) TX correlation matrix $\vm{R}_{\text{TX}}[m]$ corresponding to the maximal eigenvalue $\lambda_{\text{TX},\text{max}}[m]$. We have the length-$N_{\text{RX}}$ received vector $\vm{y}[m] = \vm{H}[m] \vm{u}_{\text{TX},\text{max}}^{\ast}[m] x[m] + \vm{n}[m]$, where $\vm{H}[m]$ are $N_{\text{RX}} \times N_{\text{TX}}$ jointly proper complex channel matrices, $x[m]$ are the transmitted signals with power $\sigma_x^2 > 0$, and the length-$N_{\text{RX}}$ vectors $\vm{n}[m]$ are zero-mean white (in time and space) jointly proper Gaussian noise vectors with known element-wise variance $\sigma_n^2 > 0$. The processes $\{\vm{H}[m]\}$, $\{x[m]\}$, and $\{\vm{n}[m]\}$ are assumed to be mutually independent. At the RX side, we process the received vector with a matched filter. The advantage of using this transmission technique lies in its simplicity since the MIMO channel reduces to a SISO channel. Moreover, the TX only requires statistical knowledge in the form of a dominant eigenvector. The (average) mismatched SNR at time instant $m$ using statistical channel knowledge from time instant $m'$ is \cite{Ispas_LQSRegions_EB, Hugl_PhD_Thesis, Viering_Validity_Spatial_Cov}
\begin{IEEEeqnarray}{rCl}
 \text{SNR}_{\text{TX}}[m,m'] &=& \frac{\vm{u}_{\text{TX},\text{max}}^H[m'] \vm{R}_{\text{TX}}[m] \vm{u}_{\text{TX},\text{max}}[m']}{\sigma_n^2 / \sigma_x^2} .
 \eqnlab{snr}
\end{IEEEeqnarray}
In the matched case, we have $m' = m$. To characterize the loss in SNR, we define the relative SNR
\begin{IEEEeqnarray}{rCl}
 \eta_{\text{SNR},\text{TX}}[m,m'] &=& \frac{\text{SNR}_\text{TX}[m,m']}{\text{SNR}_\text{TX}[m,m]} = \frac{ \vm{u}_{\text{TX},\text{max}}^H[m'] \vm{R}_{\text{TX}}[m] \vm{u}_{\text{TX},\text{max}}[m']}{\lambda_{\text{TX},\text{max}}[m]} .
 \eqnlab{snr_rel}
\end{IEEEeqnarray}
In order to analyze the non-stationarity of the spatial RX domain, we consider the reverse link with the channel $\vm{H}^T[m]$. We obtain analogously $\eta_{\text{SNR},\text{RX}}[m,m']$. Compared to the Doppler and delay domains, this technique has the advantage that the non-stationarity analysis does not require a suitable parametrization. Next, we show that the resulting measure is closely connected to the CMD.

\subsection{Algorithmic Interpretation of the CMD}
\seclab{CMD_algorithmic_interpretation}

We use the eigendecomposition of the $N_k \times N_k$ correlation matrix $\vm{R}_k[m] = \vm{U}_k[m] \vm{\Lambda}_k[m] \vm{U}_k^H[m]$ where $k$ stands for ``TX'' or ``RX''. Additionally, we define the real and non-negative eigenvalue as $\lambda_{k,l}[m] = [\vm{\Lambda}_k[m]]_{l,l}$ for $l=1,\ldots,N_k$ and the eigenvector which is the $l$th column of $\vm{U}_k[m]$ as $\vm{u}_{k,l}[m]$ . We then obtain
\begin{IEEEeqnarray}{rCl}
 \tr{\vm{R}_k[m] \vm{R}_k[m']} &=& \tr{\vm{U}_k^H[m'] \vm{R}_k[m] \vm{U}_k[m'] \vm{\Lambda}_k[m']}\IEEEnonumber\\
 &=& \sum_{l=1}^{N_k} \lambda_{k,l}[m'] \left[ \vm{U}_k^H[m'] \vm{R}_k[m] \vm{U}_k[m'] \right]_{l,l}\IEEEnonumber\\
 &=& \sum_{l=1}^{N_k} \lambda_{k,l}[m'] \vm{u}_{k,l}^H[m'] \vm{R}_k[m] \vm{u}_{k,l}[m'] .
 \eqnlab{tr_rewrite}
\end{IEEEeqnarray}
With \eqnref{tr_rewrite}, we can rewrite \eqnref{cmd} as
\begin{IEEEeqnarray}{rCl}
 \text{CMD}_k[m,m'] &=& 1 - \frac{\sum_{l=1}^{N_k} \lambda_{k,l}[m'] \vm{u}_{k,l}^H[m'] \vm{R}_k[m] \vm{u}_{k,l}[m']}{\sqrt{\sum_{l=1}^{N_k} \lambda_{k,l}^2[m] \sum_{l=1}^{N_k} \lambda_{k,l}^2[m']}} .
 \eqnlab{cmd_rewrite}
\end{IEEEeqnarray}
We can thus give an algorithmic interpretation of the CMD since \eqnref{tr_rewrite} represents the sum average signal power over individual streams weighted by the eigenvalues of the correlation matrix $\vm{R}_k[m']$. When $\vm{R}_k[m]$ and $\vm{R}_k[m']$ are both rank-one matrices, then $\eta_{\text{col},\vm{R}_k} [m,m'] = \eta_{\text{SNR},k}[m,m']$\ie the CMD is equal to one minus the relative SNR.

\section{Local Quasi-Stationarity}
\seclab{LQS_channels}

In this section, we show how to obtain the LQS regions based on the introduced measures. LQS regions are local time-frequency regions inside which the channel can be non-stationary under some restriction, hence the name quasi-stationarity. Furthermore, we consider local regions and thus we obtain the name LQS regions. We emphasize that our definition is different to the quasi-WSSUS model introduced by \cite{Bello_RTVL_Channels}. The quasi-WSSUS model can be applied to the time-frequency stationarity regions of size $(\Delta_{\tau,\text{max}} \Delta_{\nu,\text{max}})^{-1}$ discussed in \secref{DU_channels}. Inside these regions the channel statistics exhibit only minor variations and thus the channel can be assumed to be stationary. In contrast, the channel is generally non-stationary inside LQS regions, with the restriction that the non-stationarity is limited in some sense\eg by a maximal performance degradation of a selected algorithm. A visualization of the LQS regions in time and frequency in comparison to the coherence regions and the stationarity regions relevant to the quasi-WSSUS model is provided in \figref{LQS}.
\begin{figure}[!t]
 \centering
 \scalebox{0.7}{
\ifx\du\undefined
  \newlength{\du}
\fi
\setlength{\du}{15\unitlength}
\begin{tikzpicture}
\pgftransformxscale{1.000000}
\pgftransformyscale{-1.000000}
\definecolor{dialinecolor}{rgb}{0.000000, 0.000000, 0.000000}
\pgfsetstrokecolor{dialinecolor}
\definecolor{dialinecolor}{rgb}{1.000000, 1.000000, 1.000000}
\pgfsetfillcolor{dialinecolor}
\pgfsetlinewidth{0.200000\du}
\pgfsetdash{}{0pt}
\pgfsetdash{}{0pt}
\pgfsetroundjoin
{\pgfsetcornersarced{\pgfpoint{1.000000\du}{1.000000\du}}\definecolor{dialinecolor}{rgb}{1.000000, 0.341176, 0.341176}
\pgfsetfillcolor{dialinecolor}
\fill (19.400000\du,2.592893\du)--(19.400000\du,13.050000\du)--(33.707107\du,13.050000\du)--(33.707107\du,2.592893\du)--cycle;
}{\pgfsetcornersarced{\pgfpoint{1.000000\du}{1.000000\du}}\definecolor{dialinecolor}{rgb}{0.000000, 0.000000, 0.000000}
\pgfsetstrokecolor{dialinecolor}
\draw (19.400000\du,2.592893\du)--(19.400000\du,13.050000\du)--(33.707107\du,13.050000\du)--(33.707107\du,2.592893\du)--cycle;
}\pgfsetlinewidth{0.200000\du}
\pgfsetdash{}{0pt}
\pgfsetdash{}{0pt}
\pgfsetbuttcap
{
\definecolor{dialinecolor}{rgb}{0.000000, 0.000000, 0.000000}
\pgfsetfillcolor{dialinecolor}
\pgfsetarrowsend{latex}
\definecolor{dialinecolor}{rgb}{0.000000, 0.000000, 0.000000}
\pgfsetstrokecolor{dialinecolor}
\draw (17.000000\du,15.000000\du)--(16.950000\du,0.450000\du);
}
\pgfsetlinewidth{0.200000\du}
\pgfsetdash{}{0pt}
\pgfsetdash{}{0pt}
\pgfsetbuttcap
{
\definecolor{dialinecolor}{rgb}{0.000000, 0.000000, 0.000000}
\pgfsetfillcolor{dialinecolor}
\pgfsetarrowsend{latex}
\definecolor{dialinecolor}{rgb}{0.000000, 0.000000, 0.000000}
\pgfsetstrokecolor{dialinecolor}
\draw (17.000000\du,15.000000\du)--(36.200000\du,15.050000\du);
}
\definecolor{dialinecolor}{rgb}{0.000000, 0.000000, 0.000000}
\pgfsetstrokecolor{dialinecolor}
\node[anchor=west] at (32.963700\du,16.285200\du){Time};
\definecolor{dialinecolor}{rgb}{0.000000, 0.000000, 0.000000}
\pgfsetstrokecolor{dialinecolor}
\node[anchor=west] at (12.603900\du,2.290630\du){Frequency};
\pgfsetlinewidth{0.200000\du}
\pgfsetdash{}{0pt}
\pgfsetdash{}{0pt}
\pgfsetroundjoin
{\pgfsetcornersarced{\pgfpoint{1.000000\du}{1.000000\du}}\definecolor{dialinecolor}{rgb}{0.600000, 1.000000, 0.447059}
\pgfsetfillcolor{darkgreen}
\fill (22.100000\du,4.850000\du)--(22.100000\du,10.850000\du)--(31.100000\du,10.850000\du)--(31.100000\du,4.850000\du)--cycle;
}{\pgfsetcornersarced{\pgfpoint{1.000000\du}{1.000000\du}}\definecolor{dialinecolor}{rgb}{0.000000, 0.000000, 0.000000}
\pgfsetstrokecolor{dialinecolor}
\draw (22.100000\du,4.850000\du)--(22.100000\du,10.850000\du)--(31.100000\du,10.850000\du)--(31.100000\du,4.850000\du)--cycle;
}\pgfsetlinewidth{0.200000\du}
\pgfsetdash{}{0pt}
\pgfsetdash{}{0pt}
\pgfsetroundjoin
{\pgfsetcornersarced{\pgfpoint{1.000000\du}{1.000000\du}}\definecolor{dialinecolor}{rgb}{0.541176, 0.960784, 1.000000}
\pgfsetfillcolor{lightblue}
\fill (24.100000\du,6.850000\du)--(24.100000\du,8.850000\du)--(29.100000\du,8.850000\du)--(29.100000\du,6.850000\du)--cycle;
}{\pgfsetcornersarced{\pgfpoint{1.000000\du}{1.000000\du}}\definecolor{dialinecolor}{rgb}{0.000000, 0.000000, 0.000000}
\pgfsetstrokecolor{dialinecolor}
\draw (24.100000\du,6.850000\du)--(24.100000\du,8.850000\du)--(29.100000\du,8.850000\du)--(29.100000\du,6.850000\du)--cycle;
}
\definecolor{dialinecolor}{rgb}{0.000000, 0.000000, 0.000000}
\pgfsetstrokecolor{dialinecolor}
\node[anchor=west] at (24.550000\du,5.600000\du){Stationarity};
\definecolor{dialinecolor}{rgb}{0.000000, 0.000000, 0.000000}
\pgfsetstrokecolor{dialinecolor}
\node[anchor=west] at (24.700000\du,7.450000\du){Coherence};
\definecolor{dialinecolor}{rgb}{0.000000, 0.000000, 0.000000}
\pgfsetstrokecolor{dialinecolor}
\node[anchor=west] at (22.150000\du,3.305000\du){Local Quasi-Stationarity};
\end{tikzpicture}
 }
 \caption{Visualization of the time-frequency LQS regions in comparison to the coherence regions and the stationarity regions relevant to the quasi-WSSUS model.}
 \figlab{LQS}
\end{figure}
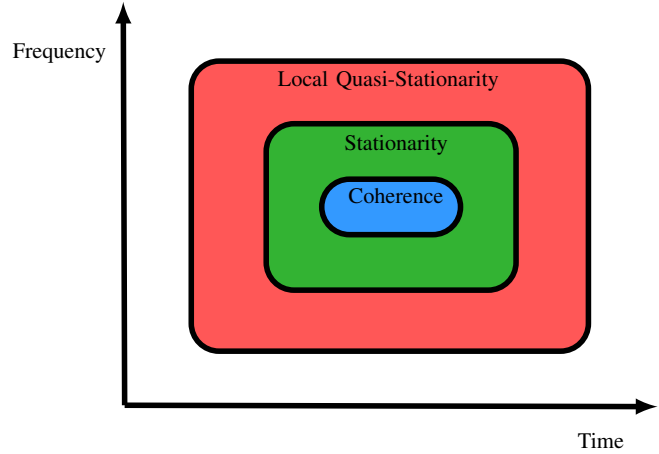

For every measure $\eta_k[m,m+\Delta_m]$, where the index $k$ indicates the considered measure, we define the average measure
\begin{IEEEeqnarray}{rCl}
 \eta_{\text{avg},k}[\Delta_m] &=& \frac{1}{|\mathcal{L}|} \sum_{m\in \mathcal{L}} \eta_k[m,m+\Delta_m]
\end{IEEEeqnarray}
with the set $\mathcal{L}$ containing all elements $m$ over which we average the measure. By using measures averaged over a measurement scenario defined by the set $\mathcal{L}$, we characterize the degree of non-stationarity of the entire scenario. This averages out effects due to specific propagation conditions\eg the change from an open environment to a street canyon. As such, we obtain the degree of non-stationarity in an average sense, which is more meaningful from an operational perspective since it is valid for the entire scenario.

We can then define the sets $\set{M}_k$ using a threshold $\eta_{\text{th}}$ as
\begin{IEEEeqnarray}{rCl}
 \set{M}_k &=& \left\{ \Delta_m \mid \eta_{\text{avg},k}[\Delta_m] > \eta_{\text{th}} \right\} .
 \eqnlab{stat_set}
\end{IEEEeqnarray}
Note that $\eta_k[m,m+\Delta_m] \in [0,1]$ holds, thus $\eta_{\text{th}}$ has to be chosen in the interval $[0,1]$. Here, the value $1$ represents stationarity and the value $0$ represents the highest degree of non-stationarity. For the algorithmic measures, this threshold immediately follows from the maximal tolerated performance degradation. These sets allow us to obtain time-independent LQS times as
\begin{IEEEeqnarray}{rCl}
 T_{\text{LQS},k} &=& \left| \set{C}_k \right| T
 \eqnlab{stat_time}
\end{IEEEeqnarray}
where $T = 1 / B_{\nu}$ denotes the spacing between the time samples, and $\set{C}_k$ is the connected subset of $\set{M}_k$ with maximum cardinality and containing the element $\Delta_m = 0$. It is important to mention that the average measures tend to exhibit a smoother behavior than the instantaneous ones. Therefore, they are better suited for a thresholding operation and consequently the definition of LQS regions.

Finally, we emphasize that we only study LQS regions and not the aforementioned stationarity regions. We thus only perform a single thresholding operation. Moreover, the choice of the threshold is clearly motivated for the algorithmic measures.

\section{Channel Measurements and Data Processing}
\seclab{measurements_and_processing}

Our MIMO channel measurement campaign \cite{Ilmenau_Reference} focused on gathering realistic channel data in an urban macrocell scenario relevant to 3GPP LTE. Channel sounding, according to the principle described in \cite{Thomae_Sounder}, was conducted at $2.53$~GHz in two bands of $45$~MHz. On the base station (BS) side, a uniform linear array (ULA) with eight antenna elements was used. Each antenna element consists of a stack of four DP patch antennas in order to form a narrow transmit beam in elevation. At the MT (passenger car), two uniform circular arrays (UCAs), one above the other, with twelve antenna elements\ie DP patch antennas, each were used. The antenna elements at the BS ULA and the MT UCA represent vertical-polarized (VP) and horizontal-polarized (HP) antennas. The BS and the MT antenna arrays are shown in \figref{antenna_arrays}. The BS served as the TX and the MT as the RX. The measurement campaign sequentially covered measurements from three BS positions with $25$~m height to 22 MT tracks. In \figref{scenario_overview}, an overview of the three MT reference tracks and the three BS positions used in this work is shown.
\tabref{measurement_campaign} summarizes the properties of the measurement campaign, see also \cite{Schneider_Channel_Reference_Data}.

\begin{figure}[!t]
 \centering
 \includegraphics[width=0.9\columnwidth]{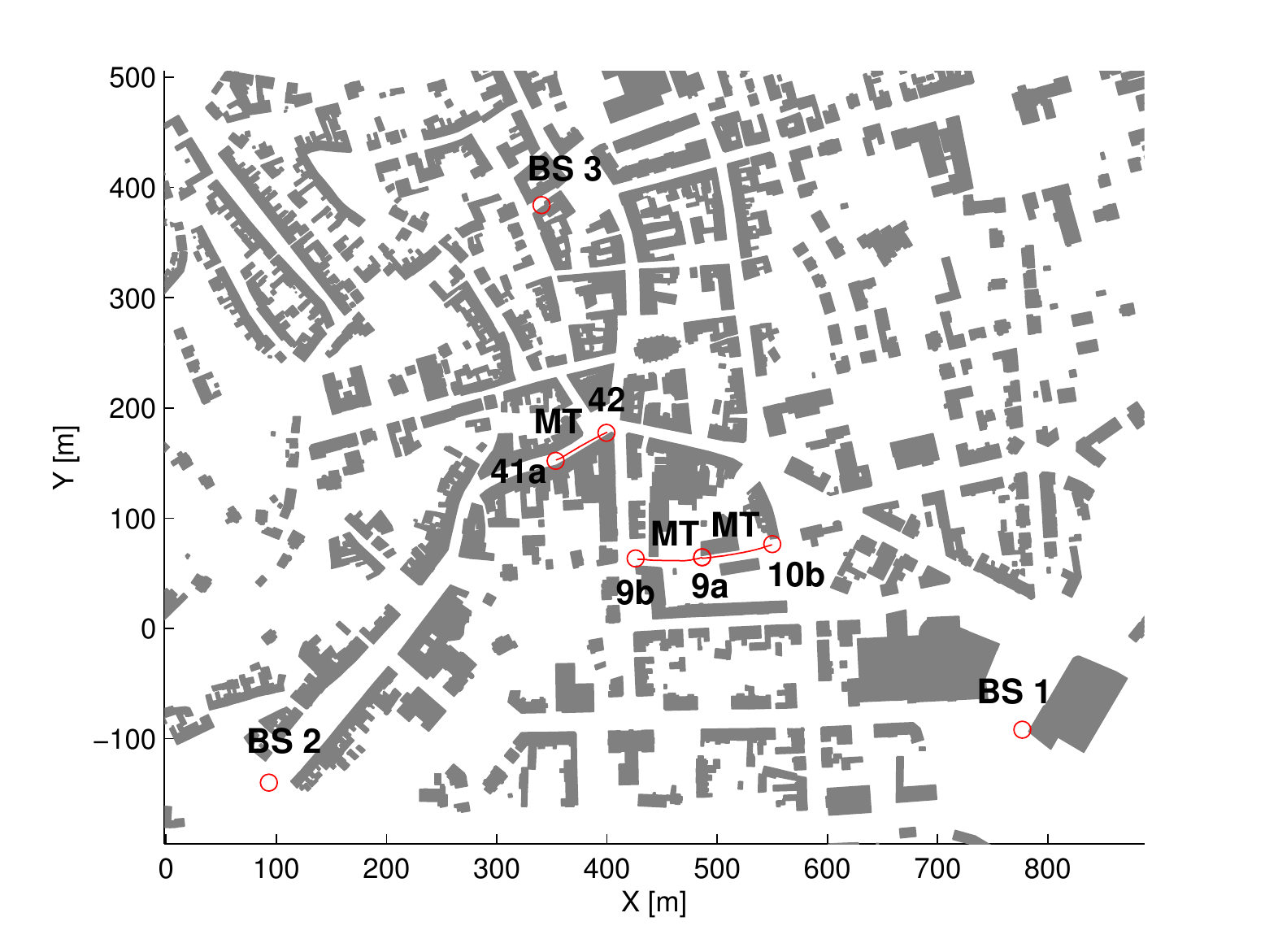}
 \caption{Overview of the MT reference tracks and the three BS positions}
 \figlab{scenario_overview}
\end{figure}

\begin{table}[!b]
\centering
\caption{Properties of the Measurement Campaign}
\begin{tabular}{l|c|c} \hline
 \mc{3}{c}{\textbf{GENERAL PROPERTIES}} \\ \hline \hline
 Scenario & \mc{2}{c}{Urban macrocell} \\ \hline
 Location & \mc{2}{c}{City center, Ilmenau, Germany} \\ \hline
 MIMO measurement setup & \mc{2}{c}{3 BSs, 22 tracks} \\ \hline
 & \mc{2}{c}{BS $1$-$2$: $680$~m} \\
 Intersite distances & \mc{2}{c}{BS $2$-$3$: $580$~m} \\
 & \mc{2}{c}{BS $3$-$1$: $640$~m} \\ \hline
 \mc{3}{c}{\textbf{CHANNEL SOUNDER PROPERTIES}} \\ \hline \hline
 Type & \mc{2}{c}{RUSK TUI-FAU, Medav GmbH} \\ \hline
 TX power & \mc{2}{c}{$46$~dBm at the power amplifier output} \\ \hline
 Center frequency $f_{\text{c}}$ & \mc{2}{c}{$2.53$~GHz} \\ \hline
 Bandwidth & \mc{2}{c}{2 bands of $45$~MHz} \\ \hline
 Time sample spacing $T_{\text{m}}$ & \mc{2}{c}{$13.1$~ms} \\ \hline
 Frequency sample spacing $F_{\text{m}}$ & \mc{2}{c}{$156.25$~kHz} \\ \hline
 MIMO sub-links & \mc{2}{c}{928 (16 BS, 58 MT antennas)} \\ \hline
 AGC switching & \mc{2}{c}{In MIMO sub-links} \\ \hline
 Positioning & \mc{2}{c}{Odometer and GPS} \\ \hline
 \mc{3}{c}{\textbf{ANTENNA PROPERTIES}} \\ \hline \hline
 & BS array & MT array \\ \hline
 Type & PULPA8 & SPUCPA 2x12 \\
 & &  + cube \\ \hline
 Height & $25$~m & $1.9$~m \\ \hline
 Beamwidth, azimuth ($3$~dB) & $100\degree$ & $360\degree$ \\ \hline
 Beamwidth, elevation ($3$~dB) & $24\degree$ & $80\degree$ \\ \hline
 Tilt & $5\degree$ down & $0$ \\ \hline
 Maximal velocity $|v_{\text{max}}|$ & $0$ & $\approx 10$~km/h \\ \hline
\end{tabular}
\tablab{measurement_campaign}
\end{table}

\begin{figure}[!t]
 \centering
 \includegraphics[width=0.4\columnwidth]{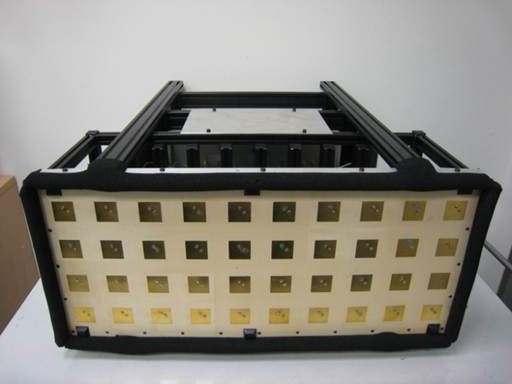}
 \includegraphics[width=0.4\columnwidth]{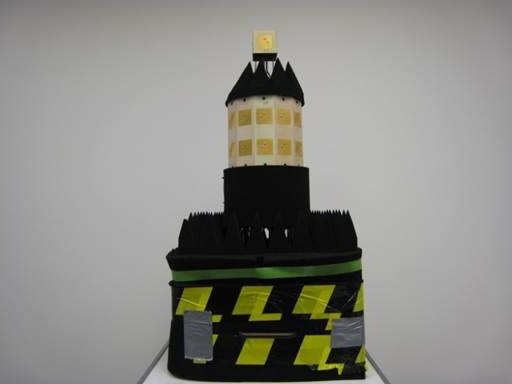}
 \caption{Antenna array at the BS (left) and at the MT (right).}
 \figlab{antenna_arrays}
\end{figure}

\subsection{Antenna Setups}
\seclab{antenna_setups}

At the BS side, we have the ULA at a height of $25$~m, and, at the MT side, we have the two UCAs that are mounted on top of each other, see \figref{antenna_arrays}. The elements chosen at the MT correspond to the front (direction of motion), the back, and the two sides of the MT.\footnote{Due to the different lengths of the connections from the antennas to the multiplexer at both the BS and the MT, additional phase shifts, different for every MIMO sub-link, are observed in the channel measurements. This effect is obviously not desired, and we make sure that it does not influence the studied measures and thus the results. We discuss this effect in the Appendix.}

\subsubsection{$4\times4$ MIMO}
\seclab{antenna_setups_4x4}

In the following, we investigate $4\times4$ MIMO with two SP and three DP antenna setups for a large BS array of length $3 \lambda_c$. The two SP antenna setups are a VP and an HP setup, and, for the DP antenna setups, we consider two co-located and one spatially separated antenna setup. For the SP antenna setups, the antenna separation at the BS is $\lambda_c$, while at the MT it is $0.5 \lambda_c$ across the two UCAs or $0.327 \lambda_c$ on the same UCA. For the two co-polarized DP antenna setups, we use two sets of co-located DP antennas at the BS and the MT; they are separated by $3 \lambda_c$ at the BS and, in one case denoted by DP-CL-1, by $0.5 \lambda_c$ across the UCAs or, in another case denoted by DP-CL-2, by $0.327 \lambda_c$ on the lower UCA at the MT. Since the neighboring antenna elements on the same UCA have a slightly different orientation (turned by 30 degrees), the VP, the HP, and the DP-CL-2 antenna setups have a wider coverage\ie opening angle, into the propagation channel (for each polarization) than the DP-CL-1 antenna setup. For the spatially separated DP antenna setup DP-SS, we use the same antenna patches as in the SP case. However, we have a separation of $2 \lambda_c$ for each polarization at the BS side. At the MT side, we only use the VP excitation on the lower UCA and only the HP excitation on the upper UCA. Note that the introduced SP and DP antenna setups result in the same array length at the BS. Additionally, we study a small BS array of length $1.5 \lambda_c$ for the same antenna setups as before with the only difference that all spacings at the BS side are divided by two. The properties of all antenna setups are summarized in \tabref{antenna_setups}.

\begin{table*}[!t]
 \centering
 \caption{Properties of the Antenna Setups with a Large/Small BS Array.}
 \begin{tabular}{cccccc} \hline
  \textbf{MIMO} & \textbf{Antenna} & \textbf{DP type} & \textbf{BS array} & \textbf{BS antenna spacing} & \textbf{MT opening} \\
  \textbf{system} & \textbf{setup} &  & \textbf{length} & \textbf{for each polarization} & \textbf{angle} \\ \hline \hline
  $4\times4$ & VP & - & $3 \lambda_c$ / $1.5 \lambda_c$ & $\lambda_c$ / $0.5 \lambda_c$ & wide \\
  $4\times4$ & HP & - & $3 \lambda_c$ / $1.5 \lambda_c$ & $\lambda_c$ / $0.5 \lambda_c$ & wide \\
  $4\times4$ & DP-CL-1 & co-located & $3 \lambda_c$ / $1.5 \lambda_c$ & $3 \lambda_c$ / $1.5 \lambda_c$ & narrow \\
  $4\times4$ & DP-CL-2 & co-located & $3 \lambda_c$ / $1.5 \lambda_c$ & $3 \lambda_c$ / $1.5 \lambda_c$ & wide \\
  $4\times4$ & DP-SS & spatially separated & $3 \lambda_c$ / $1.5 \lambda_c$ & $2 \lambda_c$ / $\lambda_c$ & wide \\ \hline
  $2\times2$ & VP-1 & - & $3 \lambda_c$ / $1.5 \lambda_c$ & $3 \lambda_c$ / $1.5 \lambda_c$ & narrow \\
  $2\times2$ & VP-2 & - & $3 \lambda_c$ / $1.5 \lambda_c$ & $3 \lambda_c$ / $1.5 \lambda_c$ & wide \\
  $2\times2$ & HP-1 & - & $3 \lambda_c$ / $1.5 \lambda_c$ & $3 \lambda_c$ / $1.5 \lambda_c$ & narrow \\
  $2\times2$ & HP-2 & - & $3 \lambda_c$ / $1.5 \lambda_c$ & $3 \lambda_c$ / $1.5 \lambda_c$ & wide \\
  $2\times2$ & DP-CL & co-located & $3 \lambda_c$ / $1.5 \lambda_c$ & - & narrow \\
  $2\times2$ & DP-SS-1 & spatially separated & $3 \lambda_c$ / $1.5 \lambda_c$ & - & narrow \\
  $2\times2$ & DP-SS-2 & spatially separated & $3 \lambda_c$ / $1.5 \lambda_c$ & - & wide \\ \hline
 \end{tabular}
 \tablab{antenna_setups}
\end{table*}

\subsubsection{$2\times2$ MIMO}
\seclab{antenna_setups_2x2}

We also investigate $2\times2$ MIMO with two VP (VP-1 and VP-2), two HP (HP-1 and HP-2), and two spatially-separated DP antenna setups (DP-SS-1 and DP-SS-2) for a large BS array of length $3 \lambda_c$. The antenna separation at the BS is $3 \lambda_c$, while at the MT it is $0.5 \lambda_c$ across the two UCAs (VP-1, HP-1, and DP-SS-1) or $0.327 \lambda_c$ on the lower UCA (VP-2, HP-2, and DP-SS-2). We again study a small BS array of length $1.5 \lambda_c$ for the same antenna setups as before with the only difference that all spacings at the BS side are divided by two. Additionally, we use a co-located DP antenna setup DP-CL for which we use the lower UCA.

\subsection{Data Processing}
\seclab{data_processing}

For the subsequent non-stationarity analysis, we use a 20 MHz band between $2.495$~GHz and $2.515$~GHz. We preprocess the data by estimating a noise level in the time-delay domain and not considering any values below it. For each time-frequency stationarity region, we normalize the channel matrices $\vm{H}[m,q]$ with a scalar factor such that the condition $\E{||\vm{h}_{\text{co}}[m,q]||_F^2} = N_{\text{co}}$ is emulated. Here, $\vm{h}_{\text{co}}[m,q]$ is a vector containing only the elements of $\vm{H}[m,q]$ corresponding to co-polarized sub-links and $N_{\text{co}}$ denotes their number. We thus effectively remove the path loss and the shadow fading, while accounting for the power loss in cross-polarized sub-links \cite{Coldrey_MIMO_DP_Channels}.

\subsubsection{Doubly Underspread Condition}

Before performing the estimation of the second-order moments of the channel, we verify that the DU condition holds. For this, we need estimates of the stationarity regions; thus, we can only perform a rough check. The maximal velocity of the MT is $|v_{\text{max}}| \approx 10$~km/h. Since the base station is fixed and we assume the scatterers to be fixed for now, we obtain a maximal Doppler shift $|\nu_{\text{max}}| = |v_{\text{max}}| f_{\text{c}} / c_0 \approx 23.4$~Hz, with the center frequency $f_{\text{c}}$ and the speed of light in vacuum $c_0$. This results in a minimal coherence time $T_{\text{coh,min}} = 1 / |\nu_{\text{max}}| = 42.7$~ms. We observe a maximal delay $\tau_{\text{max}} \approx 5~\mu$s, which gives a minimal coherence frequency $F_{\text{coh,min}} = 1 / \tau_{\text{max}} = 200$~kHz. Assuming a minimal stationarity length of $d_{\text{stat,min}} \approx 10 \lambda_c = 10 c_0/f_{\text{c}} = 1.19$~m \cite{WINNER}, a rough estimate of the minimal stationarity in time is $T_{\text{stat,min}} = 1 / \Delta_{\nu,\text{max}} = d_{\text{stat,min}} / v_{\text{max}} \approx 0.43$~s. If correlation of different delay-Doppler components is only a result of scattering from the same physical object \cite{Matz_Non-WSSUS_Channels}, this corresponds to a maximum angular spread of $\delta = 25.8\degree$ when given by $\Delta\nu_{\text{max}} = 2 \nu_{\text{max}} \sin^2 (\delta/2)$. Similarly, we estimate the minimal stationarity in frequency $F_{\text{stat,min}}$ assuming that the maximal size of an object is $w_{\text{max}} \approx 15$~m and that only components from the same object are correlated\ie $F_{\text{stat,min}} = 1 / \Delta_{\tau,\text{max}} \approx c_0 / w_{\text{max}} \approx 20$~MHz. We thus obtain $\Delta_{\tau,\text{max}} \Delta_{\nu,\text{max}} \approx 1.16\cdot10^{-7}$ and $\tau_{\text{max}} \nu_{\text{max}} \approx 1.17\cdot10^{-4}$ and thus the DU condition $\Delta_{\tau,\text{max}} \Delta_{\nu,\text{max}} \ll \tau_{\text{max}} \nu_{\text{max}} \ll 1$ is fulfilled in our scenario. With the above estimates, the minimum number of coherent samples is $3$ in time and $1$ in frequency. The minimum number of stationary samples is $32$ in time and $128$ in frequency.

\subsubsection{Estimation of the GLSF}

For DU channels, an estimation of the GLSF\ie a spectral estimator, using only a single measurement run is proposed in \cite{Matz_Doubly_Underspread_Channels, Kozek_QTFI_Estimation}. Since the channel measurements are available at discrete time and frequency instants, we are interested in a discrete representation of the GLSF. By removing the expectation operator in \eqnref{GLSF_alt}, we can obtain a GLSF estimator that is similar to the one used in \cite{Paier_Vehicular_Channel_LSF}:
\begin{IEEEeqnarray}{rCl}
 \hat{C}_{\opH}^{(\Phi)}\left(mT_{\text{m}},qF_{\text{m}};\frac{p}{B_{\Delta_t}},\frac{n}{B_{\Delta_f}}\right) &=& \sum_{s=0}^{S-1} \gamma_s \left| H^{(\op{G}_s)}[m,q;p,n] \right|^2
 \eqnlab{GLSF_est}
\end{IEEEeqnarray}
for $|p| \leq \frac{B_p-1}{2}$ and $0 \leq n \leq B_n-1$ with
\begin{IEEEeqnarray}{rCl}
 H^{(\op{G}_s)}[m,q;p,n] &=& \sqrt{T_{\text{m}} F_{\text{m}}} \sum_{m'=-\lfloor N_{\text{w,t}}/2 \rfloor}^{\lceil N_{\text{w,t}}/2 \rceil -1} \sum_{q'=-\lfloor N_{\text{w,f}}/2 \rfloor }^{\lceil N_{\text{w,f}}/2 \rceil -1} L_{\op{G}_s}^{\ast}[m',q']\IEEEnonumber\\
 &&\times L_{\opH}\left((m + m')T_{\text{m}}, (q + q')F_{\text{m}}\right) e^{-j2\pi \left( \frac{p m' T_{\text{m}}}{B_{\Delta_t}} - \frac{n q' F_{\text{m}}}{B_{\Delta_f}} \right)}
\end{IEEEeqnarray}
where $N_{\text{w,t}}$ and $N_{\text{w,f}}$ denote the window lengths in time and frequency, respectively, and $T_{\text{m}} = 1/B_{\nu}$ and $F_{\text{m}} = 1/B_{\tau}$ are the time and the frequency differences between consecutive samples, respectively. For the windows in the GLSF estimation, we use a separation into time and frequency windows\ie we have $L_{\op{G}_{(a-1)J+b}}[m,q] = u_{a}[m] v_{b}[q]$, with $a = 1,\ldots,I$, $b = 1,\ldots,J$, and $S = IJ = 1/\gamma_s$. Each window is created by a discrete prolate spheroidal sequence (DPSS)\cite{Slepian_DPSS} as proposed in \cite{Matz_Doubly_Underspread_Channels, Paier_Vehicular_Channel_LSF}. The chosen time-limited DPSSs have unit-energy and are optimally concentrated in bandwidth; they are thus a good choice for a small MSE in a DU scenario, as can be concluded from the bias-variance analysis in \cite{Matz_Doubly_Underspread_Channels}. For the window lengths in time and frequency, we use $N_{\text{w,t}} = (B_p+1)/2 = 32$ and $N_{\text{w,f}} = (B_n+1)/2 = 128$, respectively. We choose all the values inside the minimal time and frequency stationarity region to obtain a maximal amount of realizations. The time-halfbandwidth product of the DPSSs is set to $2$ in time and frequency, and the number of windows is set to $I = J = 2$ to limit the computational complexity. No claims of optimality are made for the window parameters. The Doppler and the delay PSDs are then obtained according to \eqnref{PSD_Doppler} and \eqnref{PSD_delay}, respectively.

\subsubsection{Estimation of the Correlation Matrices}

In order to obtain estimates of the correlation matrices, we approximate the ensemble averaging by an averaging in time and frequency. We average over $N_t = 16$ samples in time and over $N_f = 128$ samples in frequency. Compared to the GLSF estimation, we only take half of the samples in time since the Doppler resolution is not of concern for the spatial domains. Therefore, we can better reproduce the time-variations of the channel statistics. In total, we thus average over $2048$ ($\approx 500$ non-coherent) realizations.

\section{Results}
\seclab{results}

Using the measurement data, we evaluate the LQS distances versus the threshold for the various domains of the channel. One can thus obtain the LQS distances depending on the tolerated degree of non-stationarity. The threshold to obtain the LQS distances is applied to the averaged measures, where the averaging is performed over the whole reference scenario\ie all BSs, all MT positions on the reference tracks, and all orientations of the MT. For the MSE-based measures, we use a (nominal) SNR $\gamma = 10$~dB and a pilot spacing $L=1$. Furthermore, for the exact MSE-based measure, we set the estimation interval length N to $30$ in time and $120$ in frequency. We pick the four sub-links of the $2\times2$ DP-CL antenna setup to define the vertical-to-vertical (V-V), the horizontal-to-horizontal (H-H), the vertical-to-horizontal (V-H), and the horizontal-to-vertical (H-V) polarization combination.

\subsection{Delay and Doppler Domains}

\begin{table*}[!t]
 \centering
 \caption{LQS Distances for the Doppler and the Delay Domain with a Threshold $\eta_{\text{th}} = 0.9$.}
 \begin{tabular}{c|ccc|ccc} \hline
  \textbf{Polarization} & \mc{3}{c|}{\textbf{LQS distance [m] - Doppler}} & \mc{3}{c}{\textbf{LQS distance [m] - Delay}} \\ \cline{2-7}
  \textbf{combination} & \textbf{collinearity} & \textbf{MSE} & \textbf{MSE-ap} & \textbf{collinearity} & \textbf{MSE} & \textbf{MSE-ap} \\ \hline \hline
  V-V & 2.900 & 4.200 & 3.100 & 6.400 & 7.200 & 5.100 \\
  H-H & 2.600 & 3.500 & 2.700 & 4.400 & 5.700 & 4.100 \\
  V-H & 2.400 & 2.700 & 2.100 & 8.500 & 3.700 & 2.500 \\
  H-V & 2.300 & 2.300 & 1.800 & 6.200 & 2.700 & 1.900 \\ \hline
 \end{tabular}
 \tablab{LQS_SISO}
\end{table*}

\begin{table*}[!t]
 \centering
 \caption{Standard Deviation of the Measures for the Doppler and the Delay Domain for a Distance Offset of $-10$~m.}
 \begin{tabular}{c|ccc|ccc} \hline
  \textbf{Polarization} & \mc{3}{c|}{\textbf{Standard deviation [m] - Doppler}} & \mc{3}{c}{\textbf{Standard deviation [m] - Delay}} \\ \cline{2-7}
  \textbf{combination} & \textbf{collinearity} & \textbf{MSE} & \textbf{MSE-ap} & \textbf{collinearity} & \textbf{MSE} & \textbf{MSE-ap} \\ \hline \hline
  V-V & 0.234 & 0.169 & 0.189 & 0.139 & 0.139 & 0.160 \\
  H-H & 0.246 & 0.185 & 0.203 & 0.160 & 0.154 & 0.178 \\
  V-H & 0.211 & 0.159 & 0.169 & 0.125 & 0.131 & 0.144 \\
  H-V & 0.218 & 0.157 & 0.166 & 0.130 & 0.127 & 0.140 \\ \hline
 \end{tabular}
 \tablab{LQS_std_SISO}
\end{table*}

We first study the non-stationarity in the Doppler and the delay domain for the four polarization combinations V-V, H-H, V-H, and H-V. In \tabref{LQS_SISO}, we give the LQS distances for a threshold $\eta_{\text{th}} = 0.9$. we observe that mostly lower LQS distances are observed in the Doppler compared to the delay domain. We further find that the collinearity and the exact MSE-based measure can yield significantly different LQS distances. Moreover, the approximate MSE-based measure underestimates the LQS distances with respect to the exact MSE-based measure in all cases. Only for the MSE-based measures, we observe significantly lower LQS distances on the cross-polarized links than on the co-polarized links. Additionally, only for the collinearity and the delay domain, higher LQS distances are obtained on the cross-polarized links compared to the co-polarized links. The choice of the polarization combination has a stronger impact on the LQS distances of the delay domain than those of the Doppler domain. This can be explained by the use of directional antennas at the MT; as the main source of the Doppler shift is the movement of the MT, changing the polarization combination does not have a severe influence on the Doppler domain. Regarding the delay domain, different multipath components with different delays might be observed for each polarization combination. Next, we investigate the standard deviation of the measures for the Doppler and the delay domain for a distance offset of $-10$~m\ie from the past, in \tabref{LQS_std_SISO}. It can be seen that the standard deviations of the measures are rather high; this indicates that the measures can strongly vary over the scenario. We note that the Doppler resolution is rather limited due to the short time window used in the estimation of the GLSF. However, this is necessary to properly study the non-stationarity of the channel in time\ie to reproduce the time variations of the channel statistics.

\subsection{Spatial Domains}

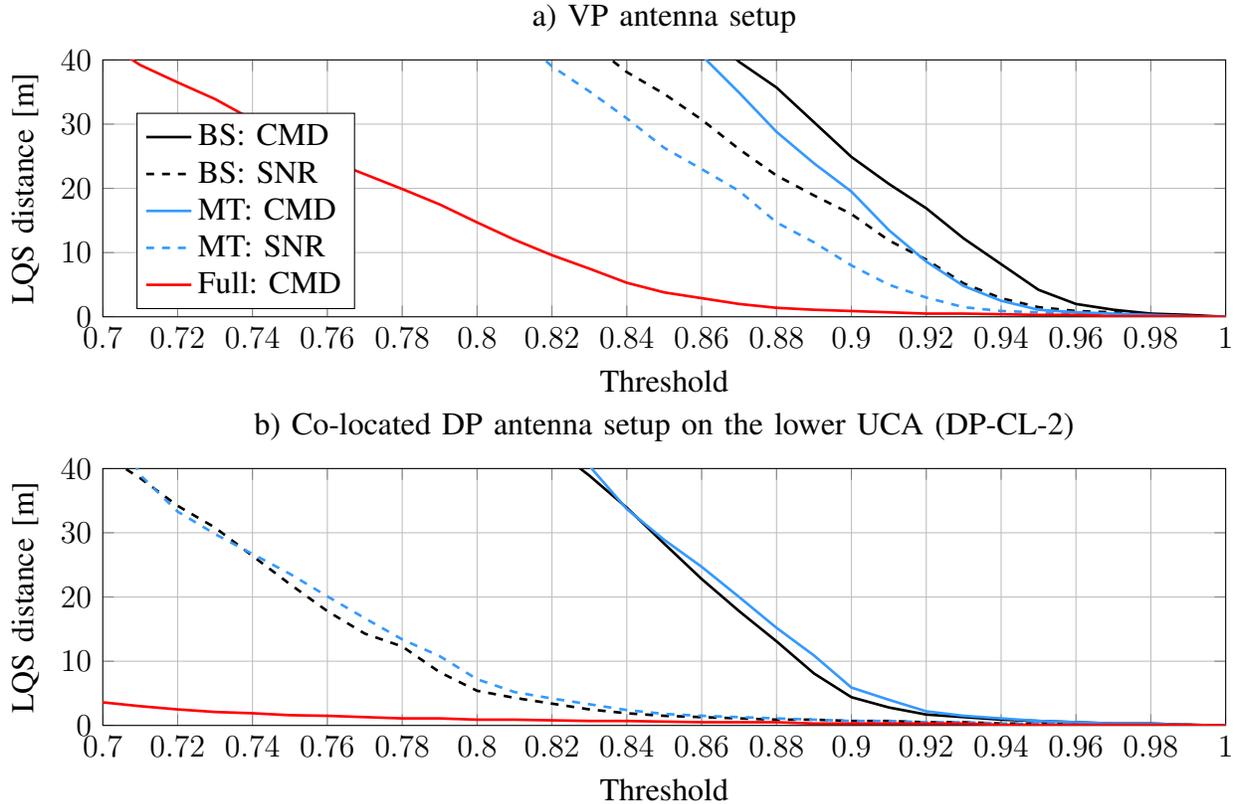
\begin{figure}[!t]
 \centering
 \begin{tikzpicture}[scale=1]
  \begin{axis}[height=5cm,width=\columnwidth,title={a) VP antenna setup}, xmin=0.7,xmax=1,ymin=0,ymax=40,xlabel={Threshold},ylabel={LQS distance [m]},grid=major,every axis legend/.append style={nodes={right}},legend pos=south west,cycle list name=mylist_LQS_MIMO]
   \addplot file {input/Stat_m_all_vs_th_CMDTX_VP_128_161_lower_averaged-16_lseg16_DPmode0.dat};
   \addplot file {input/Stat_m_all_vs_th_SNRTX_VP_128_161_lower_averaged-16_lseg16_DPmode0.dat};
   \addplot file {input/Stat_m_all_vs_th_CMDRX_VP_128_161_lower_averaged-16_lseg16_DPmode0.dat};
   \addplot file {input/Stat_m_all_vs_th_SNRRX_VP_128_161_lower_averaged-16_lseg16_DPmode0.dat};
   \addplot file {input/Stat_m_all_vs_th_CMDfull_VP_128_161_lower_averaged-16_lseg16_DPmode0.dat};
   \legend{BS: CMD, BS: SNR, MT: CMD, MT: SNR, Full: CMD}
  \end{axis}
 \end{tikzpicture}
 \begin{tikzpicture}[scale=1]
  \begin{axis}[height=5cm,width=\columnwidth,title={b) Co-located DP antenna setup on the lower UCA (DP-CL-2)}, xmin=0.7,xmax=1,ymin=0,ymax=40,xlabel={Threshold},ylabel={LQS distance [m]},grid=major,every axis legend/.append style={nodes={right}},legend pos=south west,cycle list name=mylist_LQS_MIMO]
   \addplot file {input/Stat_m_all_vs_th_CMDTX_DP-UCAl_128_161_lower_averaged-16_lseg16_DPmode1.dat};
   \addplot file {input/Stat_m_all_vs_th_SNRTX_DP-UCAl_128_161_lower_averaged-16_lseg16_DPmode1.dat};
   \addplot file {input/Stat_m_all_vs_th_CMDRX_DP-UCAl_128_161_lower_averaged-16_lseg16_DPmode1.dat};
   \addplot file {input/Stat_m_all_vs_th_SNRRX_DP-UCAl_128_161_lower_averaged-16_lseg16_DPmode1.dat};
   \addplot file {input/Stat_m_all_vs_th_CMDfull_DP-UCAl_128_161_lower_averaged-16_lseg16_DPmode1.dat};
  \end{axis}
 \end{tikzpicture}
 \caption{LQS regions vs. threshold for the spatial domain at the BS and the MT for $4\times4$ MIMO systems with a large BS array. Exemplarily, the results for the VP and the DP-CL-2 antenna setup are shown.}
 \figlab{LQS_vs_th_MIMO_4x4}
\end{figure}

\begin{table*}[!t]
 \centering
 \caption{LQS Distances for the Spatial Domain at the BS and the MT for $4\times4$ MIMO Systems with a Threshold $\eta_{\text{th}} = 0.9$.}
 \begin{tabular}{c|c|cc|cc} \hline
  \textbf{BS array} & \textbf{Antenna} & \mc{4}{c}{\textbf{LQS distance [m]}} \\ \cline{3-6}
  \textbf{size} & \textbf{setup} & \textbf{BS: CMD} & \textbf{BS: SNR} & \textbf{MT: CMD} & \textbf{MT: SNR} \\ \hline \hline
  Large & VP & 24.900 & 16.000 & 19.500 & 8.000 \\
  Large & HP & 27.000 & 19.200 & 6.900 & 2.200 \\
  Large & DP-CL-1 & 4.900 & 0.700 & 32.700 & 1.100 \\
  Large & DP-CL-2 & 4.400 & 0.700 & 5.900 & 0.700 \\
  Large & DP-SS & 6.800 & 0.900 & 4.100 & 0.700 \\ \hline
  Small & VP & $>$ 50 & $>$ 50 & 16.100 & 5.700 \\
  Small & HP & $>$ 50 & $>$ 50 & 5.200 & 1.800 \\
  Small & DP-CL-1 & 14.100 & 0.900 & 23.900 & 0.900 \\
  Small & DP-CL-2 & 13.400 & 0.900 & 4.700 & 0.700 \\
  Small & DP-SS & 25.100 & 1.100 & 3.300 & 0.600 \\ \hline
 \end{tabular}
 \tablab{LQS_MIMO_4x4}
\end{table*}

\begin{table*}[!t]
 \centering
 \caption{Standard Deviation of the Measures for the Spatial Domain at the BS and the MT for a Distance Offset of $-10$~m and $4\times4$ MIMO Systems with the Large BS Array.}
 \begin{tabular}{c|cc|cc} \hline
  \textbf{Antenna} & \mc{4}{c}{\textbf{Standard deviation of the LQS distance [m]}} \\ \cline{2-5}
  \textbf{setup} & \textbf{BS: CMD} & \textbf{BS: SNR} & \textbf{MT: CMD} & \textbf{MT: SNR} \\ \hline \hline
  VP & 0.110 & 0.164 & 0.094 & 0.148 \\
  HP & 0.107 & 0.164 & 0.106 & 0.156 \\
  DP-CL-1 & 0.097 & 0.172 & 0.064 & 0.139 \\
  DP-CL-2 & 0.097 & 0.172 & 0.092 & 0.170 \\
  DP-SS & 0.106 & 0.185 & 0.098 & 0.178 \\ \hline
 \end{tabular}
 \tablab{LQS_std_MIMO_4x4}
\end{table*}

We now investigate the non-stationarity in the spatial domain at the BS and the MT for $4\times4$ MIMO systems. In \tabref{LQS_MIMO_4x4}, we give the LQS distances with the large BS array for a threshold $\eta_{\text{th}} = 0.9$. In \figref{LQS_vs_th_MIMO_4x4}, we exemplarily depict the LQS distances vs. the treshold for the VP and the DP-CL-2 antenna setup with the large BS array of length $3 \lambda_c$, see \secref{antenna_setups}. It can be observed that, for the SP antenna setups, the LQS distances are quite high and thus the corresponding channel statistics can be reused over large distances (in an average sense). This is especially true for the spatial domain at the BS side due to its high elevation. The CMD overestimates the LQS distances with respect to the SNR-based measure; this should be considered when evaluating the update rate of the channel statistics for applications such as statistical beamforming. For the DP antenna setups, we mostly observe smaller LQS distances compared to the SP antenna setups. This is especially true when considering the SNR-based measures since, in the DP case, it is more probable for two eigenvalues to be of similar size; thus, the dominant eigenvector can easily vary along a track and the SNR-based measure yields low LQS distances. Consider now the CMD in the spatial domain at the MT of the DP-CL-1 antenna setup; here, we can observe higher LQS distances. The reason is that the DP-CL antenna setup is the only setup which does not use neighboring antenna elements at the MT; therefore, the MT can only see multipath components of the channel inside a smaller opening angle, see \secref{antenna_setups}. At the BS side, the SP antenna setups result in much higher LQS distances due to the low spacing between the individual antenna elements. In the DP-SS case, the antenna element spacing for each polarization is doubled and for the co-located DP cases it is even tripled. Thus, the DP-SS case yields lower and the co-located DP cases even lower LQS distances for the BS-related measures. Additionally, we depict the LQS distances based on the CMD of the full correlation matrix $\E{\vect{\vm{H}[m]} \left( \vect{\vm{H}[m]} \right)^H}$\ie considering the spatial domain at the BS and MT domains jointly, in \figref{LQS_vs_th_MIMO_4x4}. Obviously, the resulting LQS distances are much lower compared to the ones when studying the spatial domain at the BS and the MT individually. We further study the effect of the BS array length by evaluating the LQS distances of the small BS array of length $1.5 \lambda_c$, see \secref{antenna_setups}, in \tabref{LQS_MIMO_4x4}. It can clearly be seen that the same observations as for the larger BS array can be made with the only difference that the BS-related measures result in significantly larger LQS distances. On the MT side, we observe a small decline in the LQS distances. We also study the standard deviations of the measures for the spatial domain at the BS and the MT for a distance offset of $-10$~m for the larger BS array in \tabref{LQS_std_MIMO_4x4}. Again, we can observe strong variations of the measures over the scenario, especially for the SNR-based measures.

\begin{table*}[!t]
 \centering
 \caption{LQS Distances for the Spatial Domain at the BS and the MT for $2\times2$ MIMO Systems with a Threshold $\eta_{\text{th}} = 0.9$.}
 \begin{tabular}{c|c|cc|cc} \hline
  \textbf{BS array} & \textbf{Antenna} & \mc{4}{c}{\textbf{LQS distance [m]}} \\ \cline{3-6}
  \textbf{size} & \textbf{setup} & \textbf{BS: CMD} & \textbf{BS: SNR} & \textbf{MT: CMD} & \textbf{MT: SNR} \\ \hline \hline
  Large & VP-1 & 34.600 & 11.000 & $>$ 50 & $>$ 50 \\
  Large & VP-2 & 35.800 & 11.900 & 38.800 & 20.200 \\
  Large & HP-1 & 32.300 & 14.100 & $>$ 50 & $>$ 50 \\
  Large & HP-2 & 28.500 & 11.200 & 31.800 & 18.300 \\
  Large & DP-CL & $>$ 50 & 1.300 & $>$ 50 & 1.100 \\
  Large & DP-SS-1 & $>$ 50 & 1.100 & $>$ 50 & 1.100 \\
  Large & DP-SS-2 & $>$ 50 & 1.500 & $>$ 50 & 1.500 \\ \hline
  Small & VP-1 & 48.000 & 32.400 & $>$ 50 & $>$ 50 \\
  Small & VP-2 & 48.800 & 33.400 & 36.900 & 18.300 \\
  Small & HP-1 & 47.700 & 35.300 & $>$ 50 & $>$ 50 \\
  Small & HP-2 & 47.200 & 34.800 & 30.600 & 17.700 \\
  Small & DP-SS-1 & $>$ 50 & 1.100 & $>$ 50 & 1.100 \\
  Small & DP-SS-2 & $>$ 50 & 1.500 & $>$ 50 & 1.500 \\ \hline
 \end{tabular}
 \tablab{LQS_MIMO_2x2}
\end{table*}

In \tabref{LQS_MIMO_2x2}, we give the LQS distances for a threshold $\eta_{\text{th}} = 0.9$ of the $2\times2$ MIMO setups with the large BS array. This allows us to confirm some of the observations made in the $4\times4$ MIMO case. Consider first the SP antenna setups. By choosing the MT antennas on the same UCA (VP-2, HP-2, DP-SS-2), the MT has a wider opening angle into the propagation channel. Thus, the corresponding LQS distances for the spatial domain at the MT side are smaller than those of the antenna setups with the MT antennas on both UCAs (VP-1, HP-1, DP-SS-1), at least in the SP cases. For the DP antenna setups, we do not observe this effect since the MT has a slightly different but narrow opening angle for each polarization. Similarly, in \cite{Wallace_MIMO_Channel_Characterization}, it was found that the temporal variation of the channel is considerably decreased by using directional instead of omnidirectional antennas. As for the $4\times4$ MIMO case, the use of the small BS array results in larger LQS distances regarding the BS side, while the LQS distances characterizing the MT side slightly decrease. Note also that, in the $2\times2$ MIMO case, there is only one antenna per polarization for all DP antenna setups. Therefore, there is no antenna spacing per polarization, see \tabref{antenna_setups}, and the LQS distances with the CMDs are very high in the DP cases.

\begin{table*}[!t]
 \centering
 \caption{Correlation Coefficients Between the $4\times4$ VP MIMO with the Large BS Array and the V-V SISO Measures for a Distance Offset of $-10$~m.}
 \begin{tabular}{l||cc|cc|c|ccc|ccc} \hline
   & \mc{2}{c|}{\textbf{Sp.-BS}} & \mc{2}{c|}{\textbf{Sp.-MT}} & \textbf{Sp.-Full} & \mc{3}{c|}{\textbf{Doppler}} & \mc{3}{c}{\textbf{Delay}} \\ \cline{2-12}
  \textbf{Measure} & \textbf{CMD} & \textbf{SNR} & \textbf{CMD} & \textbf{SNR} & \textbf{CMD} & \textbf{col.} & \textbf{MSE} & \textbf{MSE-ap} & \textbf{col.} & \textbf{MSE} & \textbf{MSE-ap} \\ \hline \hline
  \textbf{Sp.-BS: CMD} & 1.000 & 0.921 & 0.396 & 0.315 & 0.822 & 0.138 & 0.125 & 0.115 & 0.286 & 0.144 & 0.106 \\
  \textbf{Sp.-BS: SNR} & 0.921 & 1.000 & 0.305 & 0.259 & 0.749 & 0.098 & 0.077 & 0.070 & 0.195 & 0.052 & 0.022 \\ \hline
  \textbf{Sp.-MT: CMD} & 0.396 & 0.305 & 1.000 & 0.911 & 0.742 & 0.230 & 0.174 & 0.159 & 0.252 & 0.199 & 0.163 \\
  \textbf{Sp.-MT: SNR} & 0.315 & 0.259 & 0.911 & 1.000 & 0.664 & 0.219 & 0.112 & 0.096 & 0.210 & 0.136 & 0.104 \\ \hline
  \textbf{Sp.-Full: CMD} & 0.822 & 0.749 & 0.742 & 0.664 & 1.000 & 0.204 & 0.158 & 0.146 & 0.258 & 0.176 & 0.138 \\ \hline
  \textbf{Dop.: col.} & 0.138 & 0.098 & 0.230 & 0.219 & 0.204 & 1.000 & 0.666 & 0.609 & 0.137 & 0.081 & 0.077 \\
  \textbf{Dop.: MSE} & 0.125 & 0.077 & 0.174 & 0.112 & 0.158 & 0.666 & 1.000 & 0.987 & 0.095 & 0.153 & 0.143 \\
  \textbf{Dop.: MSE-ap} & 0.115 & 0.070 & 0.159 & 0.096 & 0.146 & 0.609 & 0.987 & 1.000 & 0.086 & 0.149 & 0.142 \\ \hline
  \textbf{Del.: col.} & 0.286 & 0.195 & 0.252 & 0.210 & 0.258 & 0.137 & 0.095 & 0.086 & 1.000 & 0.395 & 0.376 \\
  \textbf{Del.: MSE} & 0.144 & 0.052 & 0.199 & 0.136 & 0.176 & 0.081 & 0.153 & 0.149 & 0.395 & 1.000 & 0.965 \\
  \textbf{Del.: MSE-ap} & 0.106 & 0.022 & 0.163 & 0.104 & 0.138 & 0.077 & 0.143 & 0.142 & 0.376 & 0.965 & 1.000 \\ \hline
 \end{tabular}
 \tablab{LQS_corr_4x4VP_VV}
\end{table*}

\subsection{Correlations}

We also study the correlations between the spatial measures with the $4\times4$ MIMO VP antenna setup (large BS array) and the delay and Doppler measures with the V-V polarization combination for a distance offset of $-10$~m in \tabref{LQS_corr_4x4VP_VV}. It can be observed that all correlation coefficients are positive or close to zero. The spatial measures at the BS side are mildly correlated to the spatial measures at the MT side, whereas the correlation between the Doppler and the delay measures is rather low. Moreover, the delay and Doppler measures are only slightly correlated to the spatial measures. As expected, the measures characterizing the same domain are mostly highly correlated.

\section{Conclusion}
\seclab{conclusion}

In this paper, we have presented an extensive measurement-based analysis of the non-stationarity of DP wireless channels. The measurements performed at $2.53$~GHz are representative for an urban macrocell environment and pertinent to LTE. Our approach is practically relevant since it connects the size of the LQS regions to the performance degradation of an algorithm due to outdated knowledge of the channel statistics. Additionally to our algorithmic measures, we used common measures from the literature as a basis for comparison. The analysis encompasses the Doppler, the delay, and the spatial domains of the DP channel as well as several antenna setups. The results demonstrate that LQS regions can be of significant size\ie several meters long. This motivates the reuse of channel statistics over large distances (in an average sense) for certain algorithms. We find that the polarization configuration of the BS and the MT array can have a strong impact on the LQS regions\ie the degree of non-stationarity of the channel. For example, the studied beamforming technique requires significantly higher update rates of the channel statistics in the case of DP channels. Moreover, an increase in the antenna spacing at the BS yields an increase in the degree of non-stationarity in the spatial domain at the BS. Due to the directional antennas of the UCAs at the MT, we were able to study the effects of the opening angle into the propagation channel at the MT side; it was revealed that an increase in the opening angle can substantially increase the degree of non-stationarity in the spatial domain at the MT. Overall, it is shown that the LQS distances can be strongly dependent on the chosen measure and threshold. Therefore, the choice of the measure and the threshold is crucial in assessing the LQS distances. Using the proposed algorithm-specific approach to the non-stationarity analysis, the measure and the threshold are naturally obtained from the considered algorithm and the tolerable performance degradation. The introduced approach thus provides an effective method to analyze the non-stationarity of the channel from a system perspective.


\appendix

In this appendix, we discuss the invariance of the measures to the phase offsets mentioned in \secref{antenna_setups}. It is reasonable to assume that these phase offsets are approximately time-invariant. There is, however, a frequency-dependency of the phase offsets since\eg at the TX (BS) element $k$ for $k=1,\ldots,N_{\text{TX}}$ the phase offset is $\phi_{\text{off},k}^{\text{TX}} = 2\pi d_{\text{off},k}^{\text{TX}} f_{\text{c}} / c_0 + \Delta_{\phi,\text{off},k}^{\text{TX}}$ with the shift in the phase offset $\Delta_{\phi,\text{off},k}^{\text{TX}} = 2\pi d_{\text{off},k}^{\text{TX}} \Delta_f / c_0$. Here, $d_{\text{off},k}^{\text{TX}}$ is the additional distance between a selected antenna and the multiplexer at the TX. Inserting realistic parameters as\eg $\Delta_f = 20$~MHz and $d = 0.1$~m, we obtain $\Delta_{\phi,\text{off},k}^{\text{TX}} = 0.042$, which is a relatively small difference in the phase offset. The same result holds for the phase offsets at the RX (MT) $\phi_{\text{off},k}^{\text{RX}}$ for $k=1,\ldots,N_{\text{RX}}$. In the SISO case, we can thus ignore the phase offsets when\eg estimating the GLSF. In the MIMO case, the influence of the different phase offsets is, however, not clear. We can define the channel transfer matrix including the phase offsets as
\begin{IEEEeqnarray}{rCl}
 \check{\vm{H}}[m,q] &=& \vm{D}_{\text{RX}} \vm{H}[m,q] \vm{D}_{\text{TX}}
\end{IEEEeqnarray}
where $\vm{D}_{\text{TX}}$ and $\vm{D}_{\text{RX}}$ are diagonal matrices containing the transmit and the receive phase offsets\ie  $[\vm{D}_{\text{TX}}]_{k,k} = \exp(-j \phi_{\text{off},k}^{\text{TX}})$ for $k=1,\ldots,N_{\text{TX}}$ and $[\vm{D}_{\text{RX}}]_{k,k} = \exp(-j \phi_{\text{off},k}^{\text{RX}})$ for $k=1,\ldots,N_{\text{RX}}$, respectively. We now study the influence of the phase offsets on the correlation matrices. With $\vm{D}_{\text{TX}} \vm{D}_{\text{TX}}^H = \vm{I}_{N_{\text{TX}}}$ and $\vm{D}_{\text{RX}}^H \vm{D}_{\text{RX}} = \vm{I}_{N_{\text{RX}}}$, we obtain
\begin{IEEEeqnarray}{rCl}
 \check{\vm{R}}_{\text{TX}}[m,q] &=& \E{ \check{\vm{H}}^T[m,q] \check{\vm{H}}^{\ast}[m,q] } = \vm{D}_{\text{TX}} \vm{R}_{\text{TX}}[m,q] \vm{D}_{\text{TX}}^{\ast} \\
 \check{\vm{R}}_{\text{RX}}[m,q] &=& \E{ \check{\vm{H}}[m,q] \check{\vm{H}}^H[m,q] } = \vm{D}_{\text{RX}} \vm{R}_{\text{RX}}[m,q] \vm{D}_{\text{RX}}^{\ast} .
\end{IEEEeqnarray}
It can easily be checked that the CMDs are invariant to phase offsets since we have
\begin{IEEEeqnarray}{rCl}
 \tr{\check{\vm{R}}_k[m,q] \check{\vm{R}}_k[m',q]} &=& \tr{\vm{R}_k[m,q] \vm{R}_k[m',q]}
\end{IEEEeqnarray}
where $k$ stands for ``TX'' or ``RX''. Similarly, one can show that the SNR-based measure is invariant to phase offsets.


\bibliographystyle{bib/IEEEtran}
\bibliography{bib/control,bib/IEEEabrv,bib/references}

\end{document}